\documentclass[letterpaper,aps,prl,twocolumn,superscriptaddress]{revtex4-1}

\usepackage{graphicx}
\usepackage{amssymb}
\usepackage{amsmath}
\usepackage{floatrow}
\setcitestyle{super}

\usepackage[T1]{fontenc}
\usepackage{xfrac}
\usepackage[letterpaper,margin=0.5in]{geometry}

\usepackage{hyperref}
\usepackage{fancyref}

\begin{document}
\title{Temperature tunability of quantum emitter - cavity coupling in a photonic wire microcavity with shielded sidewall loss}

\author{M. Bernard}
\address{Centre for Materials and Microsystems, Fondazione Bruno Kessler, I-38123 Povo, Italy}
\address{Deptartment of Physics, University of Trento, I-38123 Povo, Italy}
\author{M. Ghulinyan$^{1,*}$}

%

\maketitle
\label{sec:intro}
\textbf{
	Recent technological advancements have allowed to implement in solid-state cavity-based devices phenomena of quantum nature such as vacuum Rabi splitting\cite{reithmaier2004strong,yoshie2004vacuum}, controllable single photon emission\cite{pelton2002efficient,baier2004single} and quantum entanglement\cite{michler2000quantum,stevenson2006semiconductor}. For a sufficiently strong coupling between a quantum emitter and a cavity, large quality factors ($Q$) along with small modal volume ($V_{eff}$) are essential\cite{yoshie2004vacuum,hendrickson2005quantum,khitrova2006vacuum}. Here we show that by applying a 5nm Al coating to the sidewalls of a submicrometer-sized Fabry-P\'{e}rot microcavity, the cavity $Q$ can be temperature-tuned from few hundreds at room temperatures to 2$\times$10$^5$ below 30~K. This is achieved by, first, a complete shielding of the sidewall loss with ideally reflecting lateral metallic mirrors and, secondly, a dramatic decrease of the cavity's axial loss for small-sized devices due to the largely off-axis wavevector within the multilayered structure. Our findings offer a novel temperature-tunable platform to study quantum electrodynamical phenomena of emitter-cavity coupling. We demonstrate that a Rabi splitting of 2g=24~GHz (0.142~nm) can be readily achieved at 40~K in a 0.8$\mu$m-sized device, which has an $V_{eff}\approx0.0845~\mu$m$^3$, comparable to best 2D photonic crystal (PhC) nanocavities\cite{yoshie2004vacuum}.
}

The physics of strong coupling between a quantum emitter and a cavity constitutes the backbone of cavity Quantum Electrodynamics (QED)\cite{berman1994cavity}. The pioneering studies of coupling between electronic (quantum emitter) and photonic (cavity) states\cite{jaynes1963comparison,purcell1946spontaneous,carmichael1989subnatural,andreani1999strong} have triggered the expansion of cavity QED from fundamental research \cite{berman1994cavity} towards its solid-state implementation in laboratory devices\cite{hennessy2007quantum,coles2014waveguide,pelton2002efficient,reithmaier2004strong,yoshie2004vacuum,stevenson2006semiconductor,baier2004single,michler2000quantum} to quantum photonic circuits\cite{politi2008silica,o2009photonic}.
The rate $g$ at which the cavity and the emitter exchange photons is crucial to determine if the system is in weak or strong coupling regime. In the first case, following Fermi's golden rule, the free-space spontaneous emission rate of the emitter can be altered when it radiates within an electromagnetic cavity\cite{yablonovitch1987inhibited,purcell1946spontaneous}. Strong coupling manifests when the emitter and the cavity start to exchange photons coherently through a process known as vacuum Rabi oscillations\cite{khitrova2006vacuum}. This happens when $g^2\geq(\gamma_c-\gamma_e)^2/16$, where $\gamma_{c,e}$ are the cavity and emitter linewidths, respectively\cite{andreani1999strong}. The strong coupling regime has gained a particular interest since it constitutes the basis of a rich spectrum of cavity QED phenomena\cite{khitrova2006vacuum}.

The strong coupling requires several conditions to be satisfied. First, along with a narrow linewidth, $\gamma_e$, the emitter has to possess large oscillator strength $f$, in order to provide necessary coupling rates ($g\sim\sqrt f$). Secondly, the cavity needs to have large modal $Q$ ($\sim\gamma_c^{-1}$) and small volume $V_{eff}$, since $g$ scales as $V_{eff}^{-1/2}$. Among integrated solid-state cavity structures Fabry-P\'{e}rot-type (FP) micropillars\cite{pelton2002efficient,gerard1998enhanced}, PhC-slab nanocavities\cite{akahane2003high,yoshie2004vacuum,ellis2011ultralow},  microdisks\cite{mccall1992whispering,srinivasan2006cavity,pitanti2010probing}  and microspheres\cite{fan2000coupling,pelton1999ultralow} are the frequently utilized ones. Disk and sphere resonators offer Q's in excess of millions but their large mode volumes make them impractical for cavity QED studies. PhC nanocavities, which offer mode volumes as small as 0.04~$\mu$m$^3$ along with reasonably high active Q's of 20,000, require careful geometry engineering and precise technological control\cite{song2005ultra}.

\begin{figure}[b!]
	\centering
	\includegraphics[width=8.5cm]{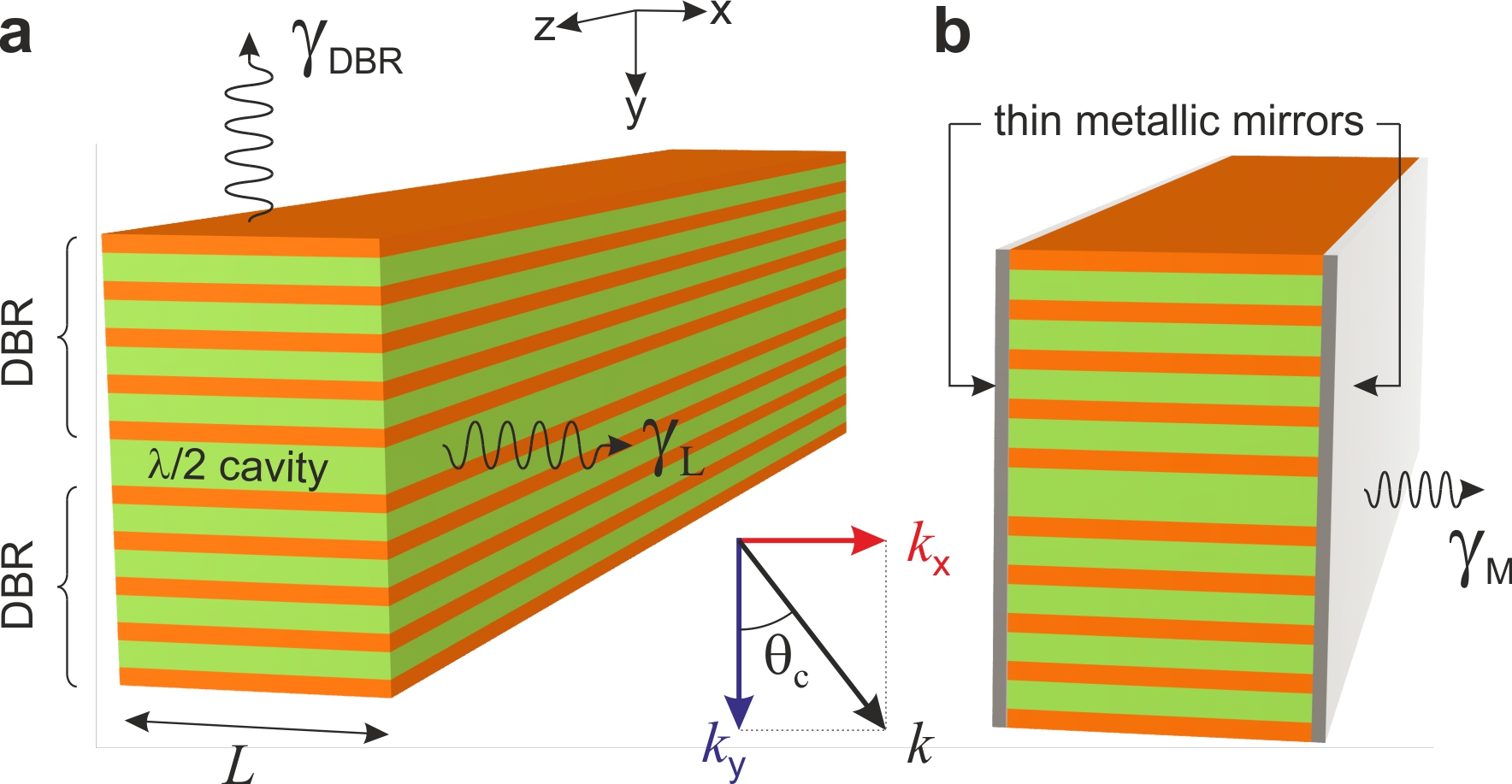}
	\caption{\textbf{Schematic representation of the multilayered FP-type photonic wire microcavity.} (a) The system loss is dominated by the dielectric mirror losses for large lateral size $L_x$ and by the lateral loss $\gamma_L$ when $L_x/\lambda\lesssim1$. (b) A sidewall Al-coverage acts as a temperature-tunable mirror, which hinders the lateral loss at low T's and boost the cavity Q by a factor of 1,000.
	}
	\label{fig:cav1}
\end{figure}

Multilayered FP microcavities, on one side, demand relatively simple and controllable technology, and, on the other side, are easily described from theoretical point of view\cite{saleh2007fundamentals}. In these devices tiny modal volumes may also be achieved when lateral dimensions squeeze down to several wavelengths. Here, however, sidewall losses due to scattering and poor mode confinement spoil the cavity Q severely \cite{reithmaier1997size,reithmaier2004strong}. Therefore, a strategy to maintain high Q's in small-sized FP microcavities is challenging and could be beneficial to cavity QED experiments.

\begin{figure*}[t!]
	\centering
	\includegraphics[width=18.5cm]{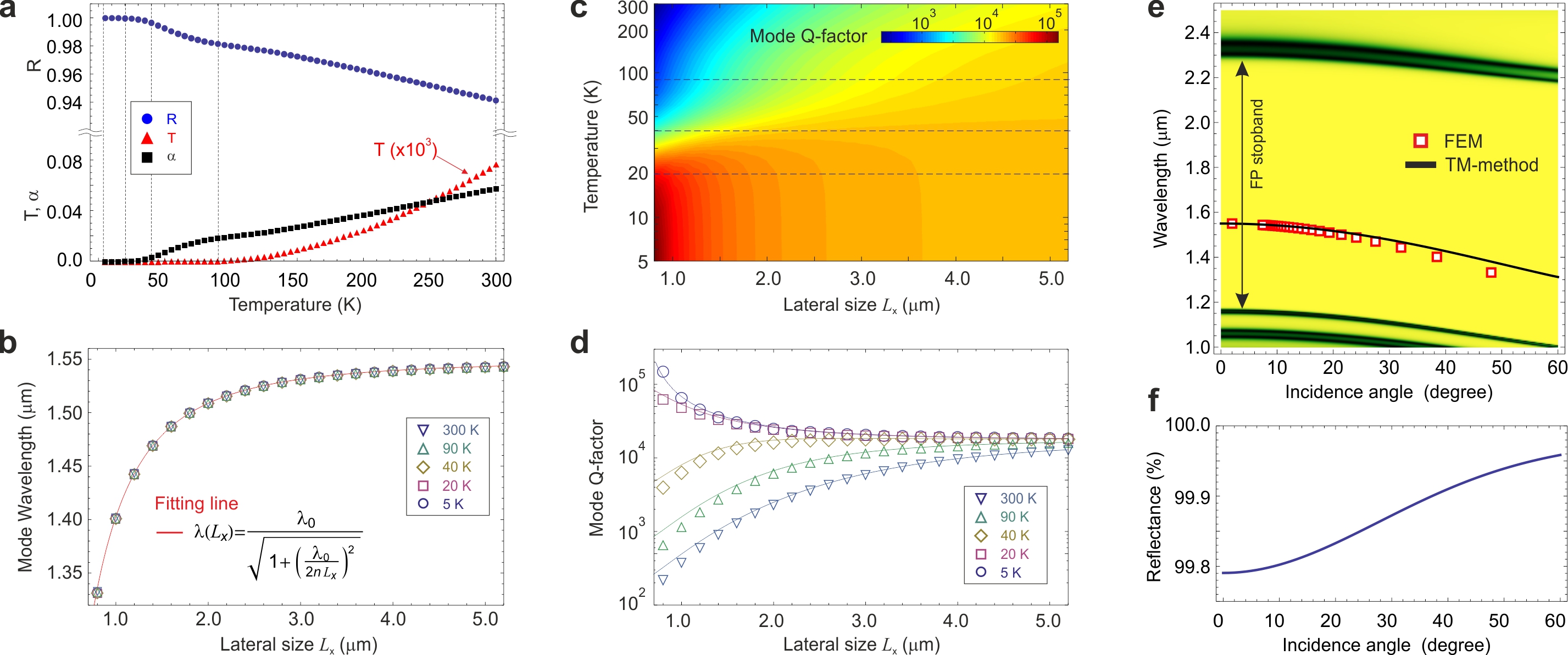}
	\caption{\textbf{Size and temperature-dependent characteristics of FP microcavities with metallic sidewall mirrors.} (a) FEM-calculated power reflection, transmission and absorption of 5~nm thick Al plate. (b) The spectral shift of the cavity resonance as a function of size is almost insensitive to temperature-dependent reflection from metallic mirrors. (c) The 2D map of calculated Q-factor (log-scale) as a function of $L_x$ and $T$ shows significant variations for small lateral size. (d) The extracted from (c) Q($L_x$)-trends for a series of selected temperatures and the corresponding analytical curves, calculated according to the model Eq.(\ref*{eq:Qfit}). (e) The power reflectance map of an infinite 1D FP microcavity under $E_z$-polarized plane-wave excitation at oblique angles has been calculated using transfer-matrix formalism. The scattered data are FEM calculation results for the FP cavity with finite lateral-size $L_x$, which is related to $\Theta_i$ through $\sfrac{\lambda_0}{2L_x}=\tan\left(\arcsin\left(\sin \Theta_i /n \right)\right)$. (f) The gradual increase of a dielectric mirror's $E_z$-wave reflectance at the FP resonance wavelength $\lambda(L_x)$ explains the cavity linewidth narrowing at larger $\Theta_i$ (smaller $L_x$).}
	\label{fig:qmap}
\end{figure*}

In the absence of material absorption, a FP microcavity is characterized by two main channels of passive loss (Fig.~\ref{fig:cav1}a): (i) the loss, $\gamma_{DBR}$, due to the finite reflectivity of dielectric mirrors and (ii) the lateral loss, $\gamma_L$, which accounts for both the electromagnetic (EM) wave scattering on the imperfect boundaries and radiation from dielectric cavity towards the environment. While the mode volume drops with decreasing the device size, $L$, the lateral loss grows exponentially. This leads to a situation where the ratio $Q/\sqrt{V_{eff}}$, and hence the coupling rate $g$ passes a maximum at an optimal device size\cite{reithmaier2004strong}.

An efficient isolation of the EM mode from the environment would be ideal to prevent the  mode Q from degradation. This, for example, can be achieved by applying highly reflecting mirrors to the sidewalls of the FP cavity (Fig.~\ref{fig:cav1}b). For this, we studied numerically a FP multilayered microcavity, which has a finite lateral dimension $L_x$ and is infinite in the third dimension (see Supplementary Methods section for details). In our approach, 5~nm thick Al coatings are applied to the sidewalls of the cavity. The lateral loss is thus substituted by the loss, $\gamma_M$, of the metallic mirror. The reflectivity of Al mirrors is temperature-tunable owing to a significant increase of the electrical conductivity, $\sigma(T)$, by five orders of magnitude when the material is cooled down from room temperatures to few kelvins (see Supplementary Fig.~S1). In particular, we find from finite-element method (FEM) calculations that the power reflection of a near-infrared wave ($\lambda=$1.55$~\mu$m) from a 5nm-thick Al sheet is R=94\% at T=300~K, while the remaining power is completely absorbed within the metal (Fig.~\ref{fig:qmap}a). The reflectance at lower temperatures grows rapidly following to a good approximation the trend $R\approx1-\left(\frac{2\epsilon_0\omega}{\pi\sigma(T)}\right)^{1/2}$, where $\omega$ is the EM wave frequency and $\epsilon_0$ is the vacuum permittivity\cite{holstein1952optical,bennett1963infrared}. The calculated $R$ is larger than 99\% already at 55~K and rapidly reaches unity (R=99.998\%) at liquid helium temperatures (T=5~K). 

The presence of metallic mirrors has imperceptible effect on the fundamental mode wavelength, $\lambda$, of the FP cavity. Our calculations show that, as expected\cite{reithmaier1997size}, the resonant wavelength scales as $\lambda(L_x)=\lambda_0/\sqrt{1+\left(\sfrac{\lambda_0}{2 n L_x}\right)^2}$, where $\lambda_0=2nd$ is the center wavelength of the laterally infinite planar cavity with mode's effective refractive index $n$ and thickness $d$. Here, for simplicity, materials dielectric functions have been considered temperature-independent. Figure \ref{fig:qmap}b plots a series of $\lambda(L_x)$-curves for lateral sizes in the range from 5.2~$\mu$m to  0.8~$\mu$m, calculated at various temperatures. It shows that the different curves are essentially coinciding, which means that, for a given size $L_x$, the cavity resonance wavelength is rather robust against temperature variation.

Quite surprisingly, we observe that in the presence of sidewall mirrors the system temperature has an unprecedented effect on the cavity Q. In a general picture, for a given lateral size, the Q grows monotonically as the temperature decreases (Fig.~\ref{fig:qmap}c). This improvement, however, is critically dependent on the cavity size (see also Supplementary Fig.~S2). In particular, while for relatively large cavities ($L_x/\lambda>3$) the change in Q amounts to a factor of two, we find an improvement of up to three orders of magnitude when the lateral size becomes smaller than the wavelength ($L_x/\lambda<1$). In fact, the Q-factor of the cavity with $L_x=0.8\mu$m (Fig.~\ref{fig:qmap}d) boosts from 200 to 200,000 when the structure is cooled down from 300K to 5K. We note that the $Q(L_x)$ dependence at room temperature is both quantitatively and qualitatively very similar for FP devices with or without metal coverage. The explanation is that in one case the Q is limited by the EM wave absorption in the metal while in the second case it is spoiled by the lateral loss from the cavity.

The remarkable result of linewidth narrowing in small-sized cavities at lower temperatures necessitates an in-depth understanding of the underlaying physics. As it appears from Figs.~\ref{fig:qmap}c,d, below T=30K the $Q(L_x)$ curves invert the decreasing trend to an increasing one gaining a factor of 1000 at the smallest size of 0.8$\mu$m. This implies that at low temperatures cavity losses decrease drastically -- a fact, which can not be explained by only close-to-unity reflectance of sidewall mirrors. To understand fully the complex mechanism underlaying this phenomenon we refer to the definition of the mode quality factor in the form $Q=\lambda(L_x)/(\gamma_{DBR}+\gamma_M)$ and analyze separately the two loss contributions. 

\begin{figure*}[t!]
	\centering
	\includegraphics[width=18.5cm]{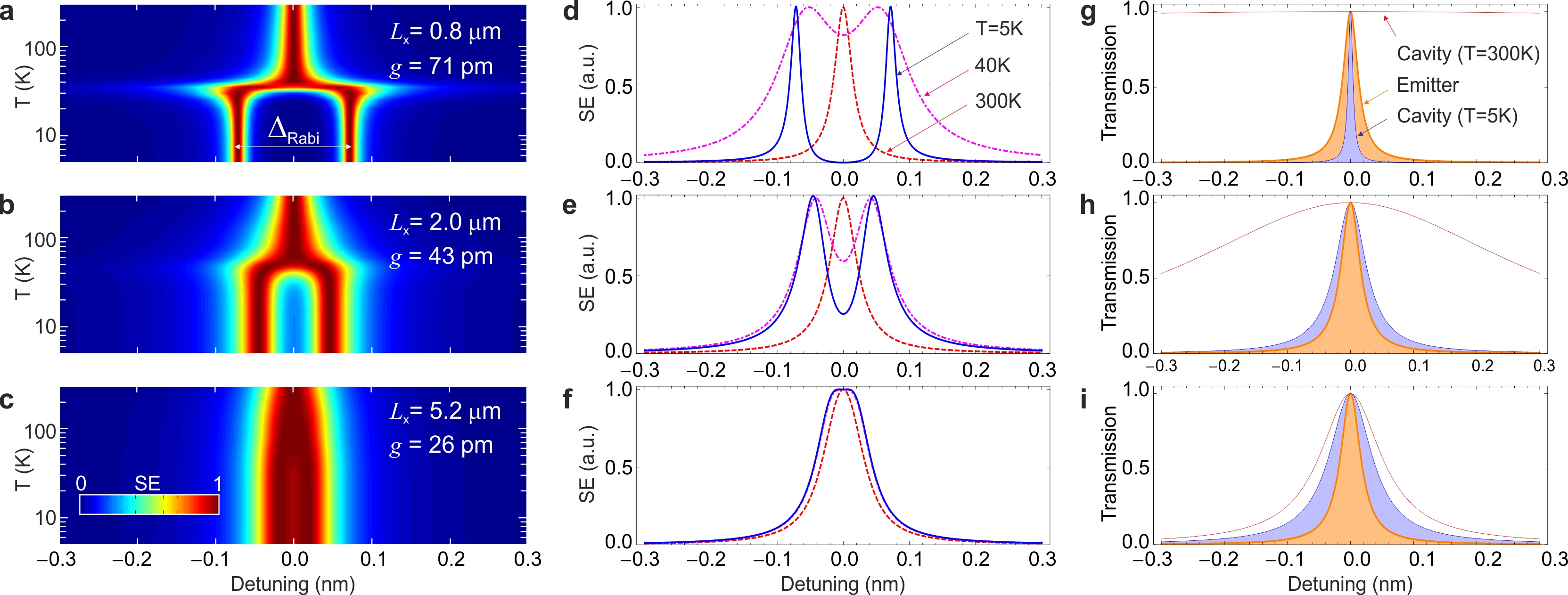}
	\caption{\textbf{QD-microcavity strong coupling and Rabi splitting.} The QD spontaneous emission map as a function of temperature and spectral detuning for photonic wire sizes (a) $L_x=0.8\mu$m, (b) 2.0~$\mu$m and (c) 5.2~$\mu$m. At a fixed oscillator strength the coupling coefficient $g$ scales with $V_{eff}^{-1/2}$ and is $71$~pm, $43$~pm and $26$~pm for different $L_x$'s, respectively. Panels (d)-(c) show the SE spectra at selected temperatures, while (g)-(i)  compare the cavity lineshape at extreme temperatures of T=5~K and 300~K to that of the emitter.}
	\label{fig:rabi}
\end{figure*}

We find that the dielectric mirror's loss $\gamma_{DBR}$ decreases as the lateral size $L_x$ shrinks. The explanation of this unexpected result is based on the fact that a normal-incident plane $E_z$-polarized wave experiences a tilted angle propagation within the multilayered dielectric structure. This angle is defined as $\Theta_c=\arctan(k_x/k_y)$, where $k_x=\pi/L_x$ and $k_y=2\pi n/\lambda_0$ are the size-quantized and axial wavevectors, respectively. According to the definition, $\Theta_c$ increases continuously as the lateral wavevector $k_x$ grows at smaller $L_x$. In terms of wave transmission, this is equivalent to the behavior of an infinite ($L_{x,y}\longrightarrow\infty$) 1D FP microcavity under an $E_z$-wave excitation at an external incidence angle $\Theta_i$, which is related to $\Theta_c$ according to Snell's law $\sin(\Theta_i)=n\sin(\Theta_c)$. Figure~\ref{fig:qmap}e shows the reflection spectrum of an infinite FP microcavity under $E_z$-polarized plane wave excitation mapped for a range of incidence angles from normal to 60$^\circ$. The cavity resonance is at $\lambda_0=1.55~\mu$m at normal incidence and drifts gradually towards shorter wavelengths as  $\Theta_i$ grows. For comparison,  FEM results for $\lambda(L_x)$ (Fig.~\ref{fig:qmap}b) are plotted in the same graph as scatter data. In this case $\Theta_i$ and $L_x$ are related according to the definition. The good agreement between these results is in strong support to our explanation. Finally, in Fig.~\ref{fig:qmap}e we show how the reflectance of the laterally infinite dielectric mirror at the resonant wavelength $\lambda(\Theta_i)$ grows progressively with the angle (decreasing size). In fact, Lorentzian fits to the resonant lines show that a higher reflection from multilayered mirrors leads to an exponential narrowing of linewidths, i.e. growing Q-factors (see Supplementary Fig.~S3). 

By considering that the metallic mirror loss $\gamma_{M}=\sfrac{-\lambda^2\ln(R(T))}{2\pi L_x}$ is cavity size\cite{saleh2007fundamentals} and temperature dependent, we model the total Q as 
\begin{equation}
Q=\frac{2\pi n}{k\gamma_{tot}}=2\pi n k\left[(\gamma_s-Ae^{-L_x/B})k_y^2-\frac{\lambda^2\ln{R(T)}}{2\pi L_x}k_x^2\right]^{-1},
\label{eq:Qfit}
\end{equation}
where the two loss contributions are weighted  for the EM field intensity in axial ($y$) and lateral ($x$) directions. Here, $\gamma_s$ is the saturation loss at very large $L_x$, while $A$ and $B$ are fitting constants. Analytical curves are shown in Fig.~\ref{fig:qmap}d as solid lines. The various trends have been excellently reproduced by plugging in Eq.(\ref*{eq:Qfit}) only the corresponding value of the Al mirror reflectance at a given T (seeFig.~\ref{fig:qmap}a).

Further calculations show that, since the cavity mode volume drops linearly with $L_x$, the ratio $Q/\sqrt{V_{eff}}$ grows monotonically as the cavity gets smaller. As noted before, this property makes the metal-coated FP cavity an excellent platform for cavity QED experiments with an added value of temperature tunability. Figure \ref{fig:rabi}a-c compares the calculated spontaneous emission (SE) maps of three cavities with lateral sizes $L_x=0.8\mu$m, 2.0~$\mu$m and 5.2~$\mu$m, respectively (see details in Supplementary Section 3). The spectra are calculated fixing the emitter linewidth\cite{hennessy2007quantum} to $\gamma_e=30$~pm ($\sim$20$~\mu$eV) and considering that the emitter and the cavity are always resonant, $\lambda_e=\lambda(L_x)$. The emitter-cavity coupling is thus defined by the $Q(T)/\sqrt{V_{eff}}$ ratio for the cavity of given size $L_x$. These results demonstrate that a whole range of quantum couplings -- from weak to strong -- can be observed within a unique device by appropriately choosing the working temperature. For example, a strong Rabi splitting of $2g=142$~pm can be readily achieved below 40~K in the smallest devices, which provide with mode volumes as small as 0.0845$~\mu$m$^3$. 

Figures \ref{fig:rabi}d-f compare the characteristic features of the SE spectra at 300~K, 40~K and 5~K. In particular, for $L_x=0.8~\mu$m size (Fig.~\ref{fig:rabi}d), the SE spectrum is essentially identical to the bare emitter's line at 300~K. This is because the cavity with Q$\sim$200 has minute effect on photon confinement and $g\ll\gamma_c/4$ leads to weak coupling (Fig.~\ref{fig:rabi}g). Contrary to this, at 5~K the cavity linewidth is much narrower than $\gamma_e$. Here, $g>\gamma_e/4$ and, thus, the strong coupling manifests as two distinctly peaked SE spectrum with a Rabi splitting that is tenfold larger than the cavity resonance width $\gamma_c$. Similar comparisons are made also for $L_x=2~\mu$m and 5.2$~\mu$m cavities in Figs.~\ref{fig:rabi}e,h and Figs.~\ref{fig:rabi}f,i, respectively. In particular, we see that for the largest cavity the emitter-cavity system remains always in weak coupling regime and no notable Rabi splitting occurs even at helium temperatures. This happens because, on one side, the larger mode volume (0.613$~\mu$m$^3$)  and, on the other, the cavity Q, which is relatively low and stable against the temperature variation, prevent the coupling rate $g$ to be larger than $|\gamma_c-\gamma_e|/4$.

Our findings have important implications on future cavity QED experiments, where the coupling between a quantum emitter and a cavity may be continuously tuned within the same semiconductor device. Whereas the sidewall loss shielding using thin metallic mirrors permits classical FP microcavities attaining very large $Q/\sqrt V$ ratios and circumvent the necessity in $Q-V$ optimization, these concepts could be also extended to waveguiding configurations paving the way towards wafer-scale integrated quantum circuitry architectures.

\subsection*{Acknowledgments}
The authors acknowledge L. Stefan and I. Carusotto for stimulating discussions. This work was partially funded by the Autonomous Province of Trento, Italy through the project ``On silicon chip quantum optics for quantum computing and secure communications -- SiQuro".


\newpage
\subsection*{Supplementary Information}

\subsection{Methods}
\subsection{The dielectric structure}
\renewcommand{\thefigure}{S\arabic{figure}}
\renewcommand{\theequation}{S\arabic{equation}}
\setcounter{figure}{0} 
\setcounter{equation}{0}  
The multilayered structure is constituted by a stack of two quarter-wavelength dielectric materials $A$ and $B$, with refractive indices $n_A=1.456$ and $n_B=3.632$ and thicknesses $d_A=\lambda_0/(4n_A)$ and $d_B=\lambda_0/(4n_B)$, respectively. The laterally infinite planar cavity has a center wavelength $\lambda_0=1550$~nm. The FP microcavity is formed by sandwiching a half-wavelength layer $AA$ between two 5.5-period distributed Bragg reflectors $BABABABAB$ (DBR). 

We notice, that any different choice for material refractive indices does not alter the findings and conclusions of this paper. In fact, all results can be reproduced for any other contrast ($n_B-n_A$) of refractive indices by appropriately choosing the number of DBR periods.

\subsection{Numerical calculations of finite size FP microcavities}

\begin{figure}[h]
	\centering
	\includegraphics[width=8.5cm]{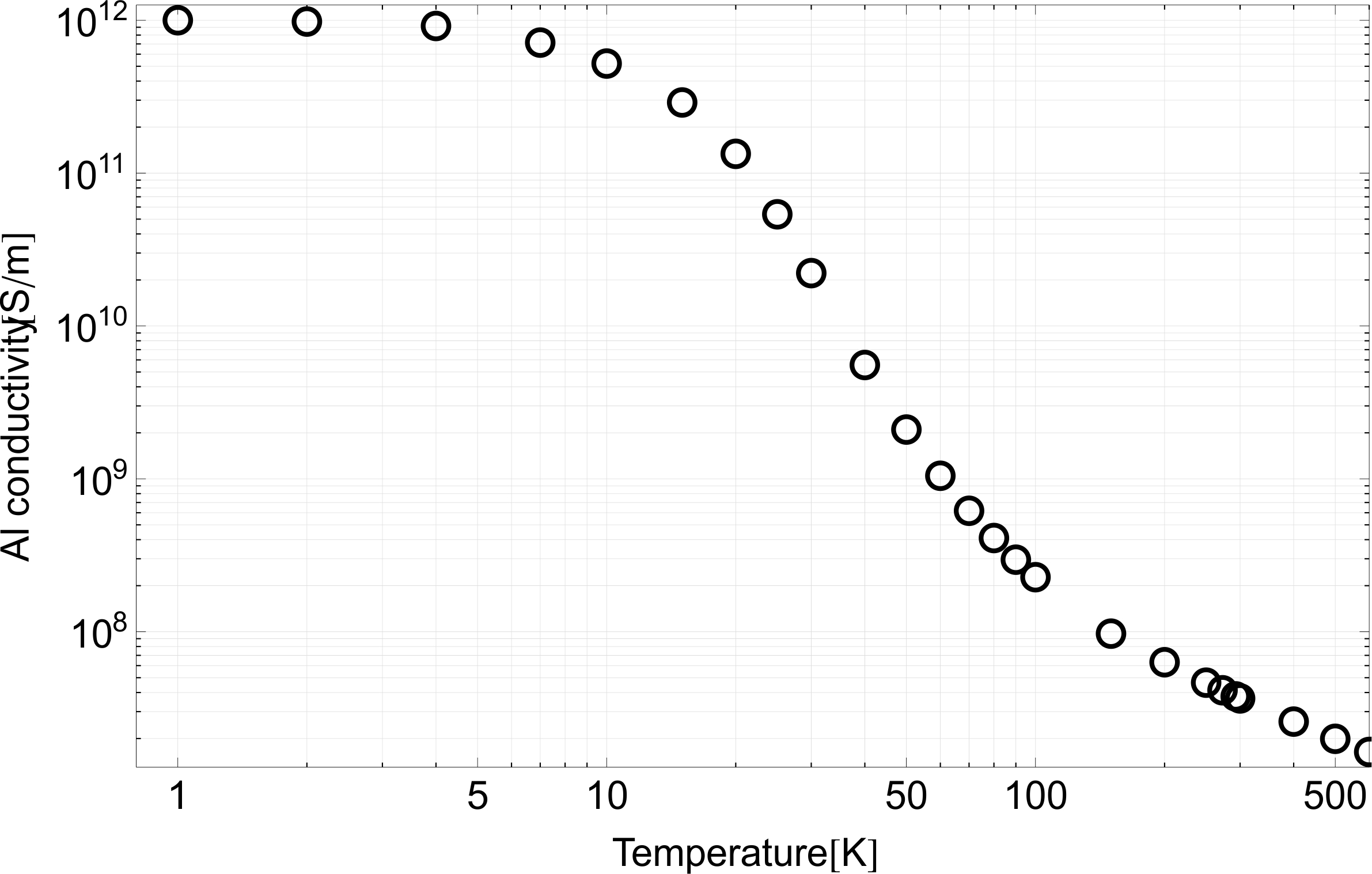}
	\caption{The conductivity vs temperature of an Al thin film (from Ref.~\cite{Alu}).}
	\label{alu}
\end{figure}

\textit{Finite-elements method} -- The resonant frequency and the cavity Q were calculated resolving the wave equation using the COMSOL$^\textrm{TM}$ commercial software's Eigenfrequency Domain Solver module. The temperature dependence of Al mirrors conductivity $\sigma(T)$ (Fig.~\ref{alu}) has been considered in the wave equation
\begin{equation}
\nabla\times(\nabla\times E)\mu_r^{-1}-k_0^2\left(\epsilon_r-\frac{\jmath\sigma(T)}{\omega\epsilon_0}\right)=0
\label{eq:WEs},
\end{equation}
where $\epsilon_r$ and $\mu_r$ are the material relative permittivity and permeability, respectively.

\begin{figure}[t!]
	\centering
	\includegraphics[width=7.5cm]{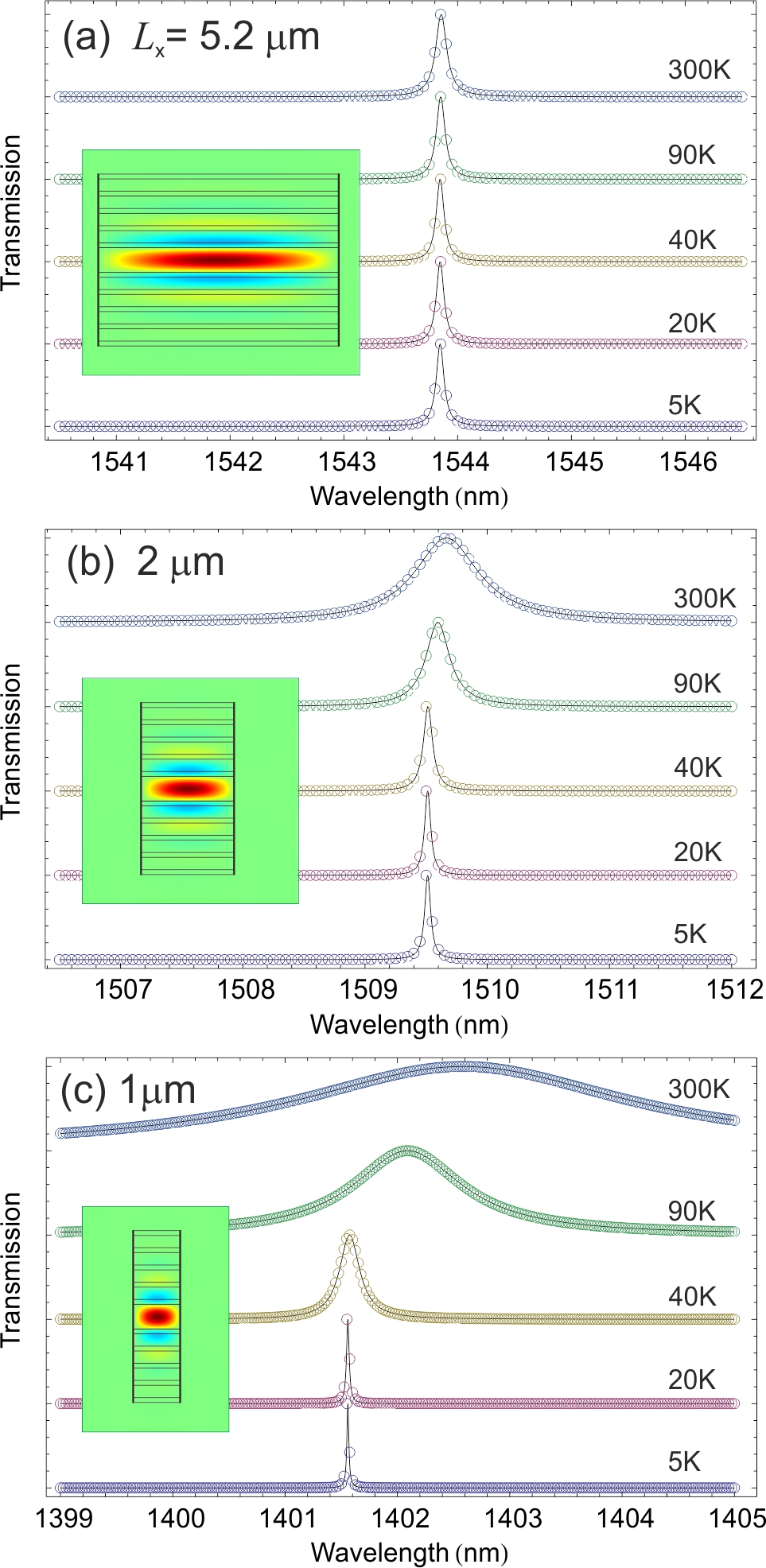}
	\caption{Calculated spectral lineshapes of cavity resonances at various temperatures for the device lateral size of (a) $L_x=5.2~\mu$m, (b) 2$~\mu$m and (c) 1$~\mu$m. Solid lines are constructed Lorentzians using Eigenfrequency domain results (center frequency and Q), while scattered data come from Frequency analysis where the plane wave transmission is calculated at free-running frequency values. Insets show the distribution of the electric field $E_z$ (normal to the screen plane) of the cavity mode.}
	\label{spectra}
\end{figure}

In Fig.~\ref{spectra}a-c the fundamental modes spectral lineshape at selected temperatures is calculated for three different cavity sizes ($L_x=5.2~\mu$m, 2$~\mu$m and 1$~\mu$m). The solid lines are based on constructed Lorentzian lineshapes following $|\frac{\Gamma/2}{\omega-\omega_0-\Gamma/2}|^2$, where the eignefrequency $\omega_0$ and the loss $\Gamma=\omega_0/Q$ were obtained from  Eigenfrequency analysis. 

The scattered data points ($\circ$) in the same graphs are the results from Frequency Domain calculations. In this approach, the power transmission of the FP microcavity is calculated by running the normal incident plane wave frequency $\omega$ around the expected resonance. An excellent agreement between the Eigenfrequency and Frequency Domain calculations is evident. The electric field distributions (mode shape and spatial extension) are shown as insets in corresponding graphs of Fig.~\ref{spectra}.

\subsection{Linewidth narrowing in planar infinite FP cavity at oblique incidence}

\textit{Transfer matrix calculations} -- The planar microcavities transmission spectra were calculated for the same layer stack $BABABABAB-AA-BABABABAB$ using a home made numeric code. 

\begin{figure}[t!]
	\centering
	\includegraphics[width=8.5cm]{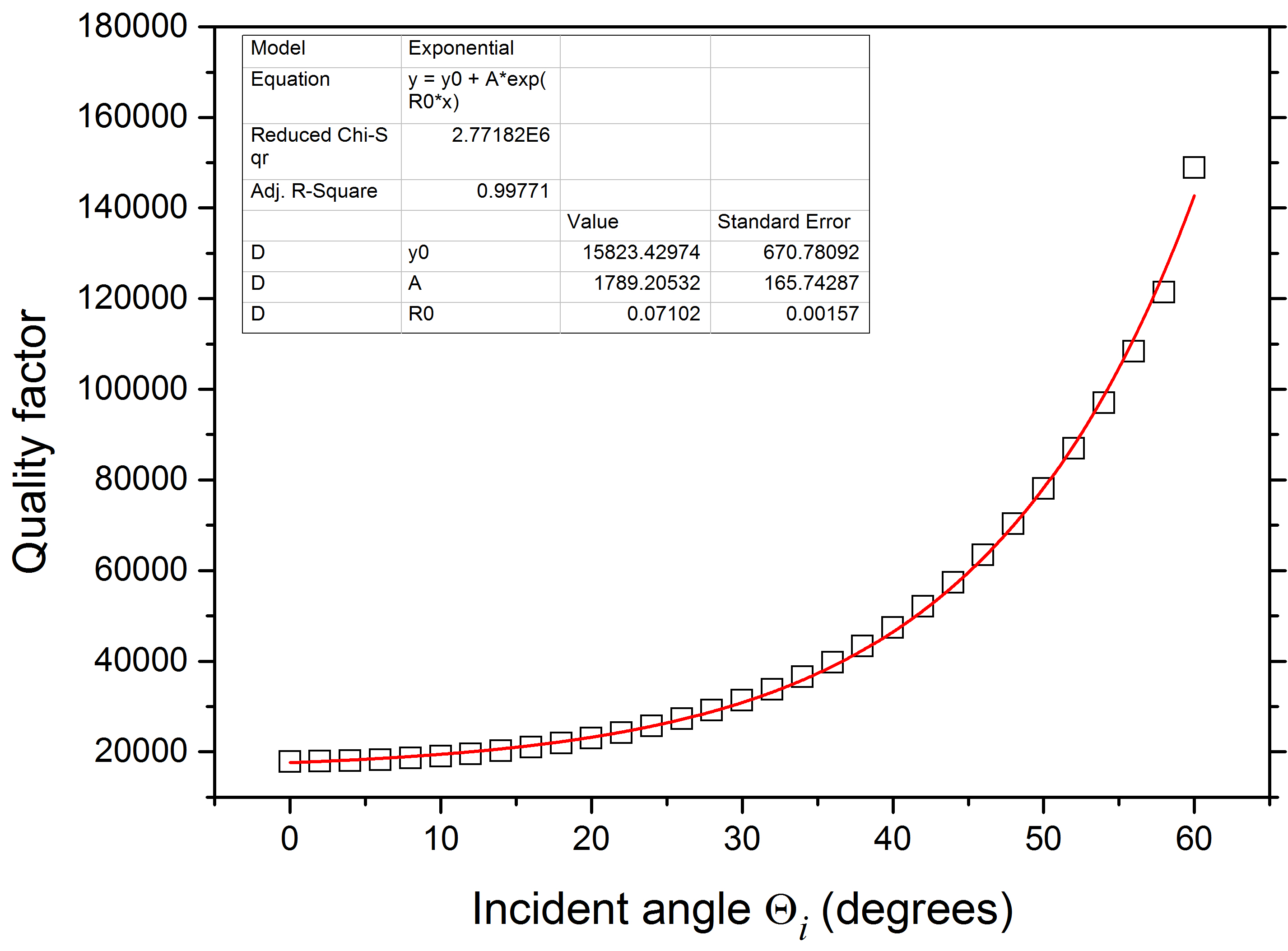}
	\caption{The transfer-matrix calculated Q of the infinite FP microcavity resonance ($\square$) against the incidence angle $\Theta_i$ and the exponential fit curve (solid line).}
	\label{quvstheta}
\end{figure}

In order to extract the cavity Q at oblique angles of incident plane TE-polarized wave, the following procedure has been adopted: first, the power transmission spectrum was calculated in a broad spectral range and, secondly,  a Lorentzian fit was applied to the resonant transmission line. Finally, the extracted Q's were plotted against the incident angle and fitted using an exponential function (Fig.~\ref{quvstheta}).

\subsection{Spontaneous Emission spectrum calculations}
We have calculated the Spontaneous Emission (SE) spectra using Eqs.~9,10 from Ref.~\cite{andreani1999strong}:
\begin{equation}
SE(\omega,T)\sim\left|\frac{\Omega_+ -\omega_0+\iota\frac{\gamma_c}{2}}{\omega-\Omega_+}-\frac{\Omega_- -\omega_0+\iota\frac{\gamma_c}{2}}{\omega-\Omega_-}\right|^2,
\end{equation}

\begin{equation}
\Omega_{\pm}(T)=\omega_0\pm\sqrt{g^2-\left(\frac{\gamma_c-\gamma_{em}}{4}\right)^2}-\iota\frac{\gamma_c+\gamma_{em}}{4}
\label{eq:RabiFreqs}
\end{equation}
where $\omega_0$ is the center frequency of the cavity resonance, $\gamma_c$ and $\gamma_{em}$ are the cavity and emitter linewidths, respectively, $g$ is the coupling rate. In strong coupling conditions, when $g^2\geq(\gamma_c-\gamma_e)^2/16$ and $\Omega_{\pm}$ describe the doublets of Lorentzians separated by the Rabi frequency 2$g$ and individual effective linewidths of $\sfrac{(\gamma_c+\gamma_{em})}{4}$.

The SE spectra are strongly dependent on the system temperature via $\omega_0(T)$ and  $\gamma_c(T)$.


\subsection*{References}

\begin{thebibliography}{99}
	
	\bibitem{reithmaier2004strong}
	Reithmaier, J.~P. \textit{et al}. Strong coupling in a single quantum dot--semiconductor microcavity system.
	\textit{Nature} \textbf{432,} 197--200 (2004).
	
	\bibitem{yoshie2004vacuum}
	Yoshie, T. \textit{et al}. Vacuum rabi splitting with a single quantum dot in a photonic crystal nanocavity.
	\textit{Nature} \textbf{432,} 200--203 (2004).
	
	\bibitem{pelton2002efficient}
	Pelton, M. \textit{et al}. Efficient source of single photons: A single quantum dot in a micropost microcavity. 
	\textit{Phys. Rev. Lett}. \textbf{89,} 233602 (2002).
	
	\bibitem{baier2004single}
	Baier, M.~H. \textit{et al}. Single photon emission from site-controlled pyramidal quantum dots.
	\textit{Appl. Phys. Lett.} \textbf{84,} 648--650 (2004).
	
	\bibitem{michler2000quantum}
	Michler, P. \textit{et al}. A quantum dot single-photon turnstile device.
	\textit{Science} \textbf{290,} 2282--2285 (2000).
	
	\bibitem{stevenson2006semiconductor}
	Stevenson, R.~M. \textit{et al}. A semiconductor source of triggered entangled photon pairs.
	\textit{Nature} \textbf{439,} 179--182 (2006).
	
	\bibitem{hendrickson2005quantum}
	Hendrickson, J. \textit{et al}. Quantum dot photonic-crystal-slab nanocavities: Quality factors and	lasing.
	\textit{Phys. Rev. B} \textbf{72,} 193303 (2005).
	
	\bibitem{khitrova2006vacuum}
	Khitrova, G., Gibbs, H.~M., Kira, M., Koch, S.~W., \& Scherer, A.
	Vacuum rabi splitting in semiconductors.
	\textit{Nature Phys.} \textbf{2,} 81--90 (2006).
	
	\bibitem{berman1994cavity}
	Berman, P.~R. \textit{Cavity Quantum Electrodynamics}. (Academic Press, Inc., Boston, 1994)
	
	\bibitem{jaynes1963comparison}
	Jaynes, E.~T. \& Cummings, F.~W.
	Comparison of quantum and semiclassical radiation theories with
	application to the beam maser.
	\textit{Proc. IEEE} \textbf{51,} 89--109 (1963).
	
	\bibitem{purcell1946spontaneous}
	Purcell, E.~M. Spontaneous emission probabilities at radio frequencies.
	\textit{Phys. Rev.} \textbf{69,} 681 (1946).
	
	\bibitem{carmichael1989subnatural}
	Carmichael, H.~J., Brecha, R.~J., Raizen, M.~G., Kimble, H.~J. \& Rice, P.~R.
	Subnatural linewidth averaging for coupled atomic and cavity-mode	oscillators.
	\textit{Phys. Rev. A} \textbf{40,} 5516--5519 (1989).
	
	\bibitem{andreani1999strong}
	Andreani, L.~C., Panzarini, G. \& G{\'e}rard, J.-M. Strong-coupling regime for quantum boxes in pillar microcavities: Theory.
	\textit{Phys. Rev. B} \textbf{60,} 13276--13279 (1999).
	
	\bibitem{hennessy2007quantum}
	Hennessy, K. \textit{et al}. Quantum nature of a strongly coupled single quantum dot--cavity
	system. \textit{Nature} \textbf{445,} 896--899 (2007).
	
	\bibitem{coles2014waveguide}
	R.~J. \textit{et al}. Waveguide-coupled photonic crystal cavity for quantum dot spin	readout.
	\textit{Opt. Express} \textbf{22,} 2376--2385 (2014).
	
	\bibitem{politi2008silica}
	Politi, A., Cryan, M.~J., Rarity, J.~G., Yu, S. \& and O'Brien, J.~L.
	Silica-on-silicon waveguide quantum circuits.
	\textit{Science} \textbf{320,} 646--649 (2008).
	
	\bibitem{o2009photonic}
	O'Brien, J.~L., Furusawa, A. \& Vu{\v{c}}kovi{\'c}, J. Photonic quantum technologies. \textit{Nature Photon.} \textbf{3,} 687--695 (2009).
	
	\bibitem{yablonovitch1987inhibited}
	Yablonovitch, E. Inhibited spontaneous emission in solid-state physics and	electronics.
	\textit{Phys. Rev. Lett.} \textbf{58,} 2059--2062 (1987).
	
	\bibitem{gerard1998enhanced}
	G\'{e}rard, J.-M. \textit{et al}.
	Enhanced spontaneous emission by quantum boxes in a monolithic	optical microcavity.
	\textit{Phys. Rev. Lett.} \textbf{81,} 1110--1113 (1998).
	
	\bibitem{akahane2003high}
	Akahane, Y., Asano, T., Song, B.-S. \& Noda, S.
	High-Q photonic nanocavity in a two-dimensional photonic crystal.
	\textit{Nature} \textbf{425,} 944--947 (2003).
	
	\bibitem{ellis2011ultralow}
	Ellis, B. \textit{et al}. Ultralow-threshold electrically pumped quantum-dot photonic-crystal
	nanocavity laser.
	\textit{Nature Photon.} \textbf{5,} 297--300 (2011).
	
	\bibitem{mccall1992whispering}
	McCall, S.~L., Levi, A.~F.~J.,Slusher, R.~E., Pearton, S.~J. \& Logan, R.~A.
	Whispering-gallery mode microdisk lasers.
	\textit{Appl. Phys. Lett.} \textbf{60,} 289--291 (1992).
	
	\bibitem{srinivasan2006cavity}
	Srinivasan, K., Borselli, M., Painter, O., Stintz, A. \& Krishna, S. 
	Cavity Q, mode volume, and lasing threshold in small diameter AlGaAs microdisks with embedded quantum dots.
	\textit{Opt. Express} \textbf{14,} 1094--1105 (2006).
	
	\bibitem{pitanti2010probing}
	Pitanti, A., Ghulinyan, M., Navarro-Urrios, D.,  Pucker, G. \& Pavesi, L.
	Probing the spontaneous emission dynamics in Si-nanocrystals-based microdisk resonators.
	\textit{Phys. Rev. Lett.}  \textbf{104,} 103901, (2010).
	
	\bibitem{fan2000coupling}
	Fan, X., Palinginis, P., Lacey, S., Wang, H. \& Lonergan, M.~C.
	Coupling semiconductor nanocrystals to a fused-silica microsphere: a	quantum-dot microcavity with extremely high Q factors.
	\textit{Opt. Letters} \textbf{25,} 1600--1602 (2000).
	
	\bibitem{pelton1999ultralow}
	Pelton, M. \& Yamamoto, Y.
	Ultralow threshold laser using a single quantum dot and a microsphere cavity.
	\textit{Phys. Rev. A} \textbf{59,} 2418--2421 (1999).
	
	\bibitem{song2005ultra}
	Song, B.-S., Noda, S., Asano, T. \& Akahane, Y.
	Ultra-high-Q photonic double-heterostructure nanocavity. 
	\textit{Nature Mater.} \textbf{4,} 207--210 (2005).
	
	\bibitem{saleh2007fundamentals}
	Saleh, B.~E.~A. \& Teich, M.~C.
	\textit{Fundamentals of Photonics: Second Edition}.
	(John Wiley \& Son, Hoboken, 2007).
	
	\bibitem{reithmaier1997size}
	Reithmaier, J.~P. \textit{et al}. 
	Size dependence of confined optical modes in photonic quantum dots.
	\textit{Phys. Rev. Lett.} \textbf{78,} 378--381 (1997).
	
	\bibitem{holstein1952optical}
	Holstein, T.
	Optical and infrared reflectivity of metals at low temperatures.
	\textit{Phys. Rev.} \textbf{88,} 1427 (1952).
	
	\bibitem{bennett1963infrared}
	Bennett, H.~E., Silver, M. \& Ashley, E.~J.
	Infrared reflectance of aluminum evaporated in ultra-high vacuum.
	\textit{JOSA} \textbf{53,} 1089--1095 (1963).
	
	\bibitem{Alu}
	Desai, P.~D., James, H.~M. and Ho, Ch.~Y. Electrical resistivity of aluminum and manganese.
	\textit{J. Phys. Chem. Ref. Data} \textbf{13,} 1131--1172 (1984).
	
\end{thebibliography}
\end{document}